# Adaptation dynamique de services

**\*Marcel Cremene, \*\*Michel Riveill, \*\*\*Christian Martel, \*Calin Loghin, \*Costin Miron**

*\* Université Technique Cluj-Napoca, Roumanie,*
*cremene@com.utcluj.ro*
*\*\* Laboratoire I3S, Sophia Antipolis, Université de Nice, France,*
*riveill@essi.fr*
*\*\*\* Equipe SysCom, Université de Savoie, Bourget-du-Lac, France,*
*Christian.Martel@univ-savoie.fr*

*RÉSUMÉ. Cet article propose une architecture logicielle qui rend possible l'adaptation dynamique de services construits par assemblage de composants, en fonction de contextes d'utilisation variés. Dans le cadre de cette première expérimentation le contexte concerne les besoins des utilisateurs. La particularité principale de notre proposition est que le comportement de chaque service par rapport à son contexte d'utilisation est évalué à partir de l'analyse du comportement de chaque composant constituant le service. Pour cela, nous utilisons des profils qui décrivent non seulement les éléments du contexte mais aussi chaque composant constituant le service. Un Adaptateur analyse la conformité entre les différents profils de chaque composant et les profils des éléments du contexte. L'Adaptateur détecte les points d'inadaptation, cherche et applique aux différents composants du service les modifications nécessaires pour rétablir cette compatibilité par modification de paramètres de configuration, par ajout, retrait ou remplacement de composants.*

*ABSTRACT. This paper proposes a software architecture for dynamical service adaptation. The services are constituted by reusable software components. The adaptation's goal is to optimize the service function of their execution context. For a first step, the context will take into account just the user needs but other elements will be added. A particular feature in our proposition is the profiles that are used not only to describe the context's elements but also the components itself. An Adapter analyzes the compatibility between all these profiles and detects the points where the profiles are not compatibles. The same Adapter search and apply the possible adaptation solutions: component customization, insertion, extraction or replacement.*

*MOTS-CLÉS : Service adaptatif, Composant logiciel, Architecture d'applications*

## 1. Introduction

Le fonctionnement d'une application de services doit tenir compte des différents éléments qui interagissent avec son fonctionnement :

–   Une application de service est généralement construite par assemblage d'une grande variété de *composants logiciels*. Ces briques logicielles préexistantes sont réutilisées pour construire l'application. Produites en continue par des différents constructeurs, elles sont adaptées au contexte précis de leur utilisation. L'Internet devient ainsi un marché libre et dynamique de composants réutilisables.

–   Les usages de ces applications utilisent divers *types de terminaux*. Ils apparaissent en continu sur le marché et il est impossible de prévoir à priori l'ensemble des plates-formes cibles en particulier dans le domaine des terminaux mobiles. Nous faisons l'hypothèse que tous possèdent une capacité de calcul et de la mémoire RAM en quantité suffisante, et divers moyens de connexion (GSM, GPRS, Lan, 802.11 ou bluetooth).

–   Les *besoins des utilisateurs* sont divers et évoluent en continu. Certains de ceux-ci dépendent du contexte physique (position géographique, bruit externe) ou du contexte social (présence d'autres personnes). Par exemple, il peut être nécessaire de compléter un service pour l'adapter aux spécificités de l'utilisateur : déficient visuel ayant besoin d'une IHM adaptée, enfant nécessitant un vocabulaire simple, locuteur anglais ne sachant lire que l'anglais.

Nous regroupons l'ensemble de ces éléments énumérés dans la notion de *contexte* du service. Ce contexte est variable et il influence en permanence les services selon différents points de vue. L'adaptation de services au contexte est un problème important dont la solution est variable dans le temps.

Ayant fait le choix de considérer qu'une application de service est construite par assemblage de composants, nous considérons que l'adaptation de celle-ci s'effectue au niveau de son architecture par ajout/retrait/remplacement ou para métrisation de ses composants. Nous avons testé notre proposition dans le cadre des services du cartable électronique déployés à l'Université de Savoie.

Cet article est organisé de la manière suivante : la section 2 est constitué d'un état de l'art du domaine et présente quelques modèles de composants et quelques architectures pour l'adaptation de services. A partir d'une étude des limitations des solutions existantes, nous présentons par l'intermédiaire d'un exemple illustratif notre proposition dans la section 3. La section 4 décrit le prototype qui nous avons implémenté validant l'architecture. La conclusion présente les principaux résultats obtenus mais aussi les faiblesses de notre proposition.

## 2. L'existant

### 2.1. *Modèles de composants réutilisables*

Un premier type de composant qui attire notre attention est le composant logiciel « compact ». Différents modèles industriels existent, certains comme le modèle COM, .Net ou EJB possèdent un faible pouvoir de reconfiguration. D'autres comme CCM - Corba Component Model, [OMG 02], Java Beans ou Fractal [BRU 02] peuvent permettre une reconfiguration dynamique plus aisées de l'assemblage précédemment réalisé. Le modèle Fractal permet une composition hiérarchisée. Un composant est généralement décrit par un langage de type IDL. Dans [FUJ 04] les auteurs proposent une description sémantique pour les composants, basée sur des ontologies et des graphes sémantiques. Il est possible de produire sous la forme d'un ADL - langage de description d'architecture [FAL 03], la manière dont les différents composants sont assemblés pour construire un service.

Un modèle de composant émerge dans le monde industriel : les services Web. Ils sont accessibles par le protocole de communication SOAP, décrits à l'aide d'un IDL particulier (WSDL) et implémentés dans un langage de programmation. La description de l'interface d'un service Web peut être complétée avec le langage DAML-S [ANK 02] qui introduit des éléments de description sémantique : pré-conditions, effets collatéraux, modèle interne du service sous forme d'un processus. Il est possible d'assembler des services Web grâce à des langages « d'orchestration » comme BPEL4WS [PEL 03] par exemple. Dans le cas de services Web, l'assemblage a une dimension temporelle et est réalisé sous la forme d'un workflow.

### 2.2. *La description du contexte*

Le contexte d'une service est composé de l'utilisateur, des ressources nécessaire à son exécution et les divers éléments qui influent son fonctionnement. Il existe plusieurs formats émergents basés sur RDF pour représenter les divers éléments du contexte. Dans sa thèse [LEM 04] utilise le cadre CC/PP - Composite Capabilities/Preferences Profiles, [W3C 04] pour décrire les caractéristiques des terminaux mais aussi les préférences utilisateurs. Le standard MPEG21 contient aussi une partie pour la description des caractéristiques liées aux réseaux et aux terminaux : formats supportés, capacité d'affichage vidéo et audio, capacité de stockage, périphériques, etc. Dans l'article [ROS 04], chaque élément du contexte physique : salle de cours, matériaux du cours, professeurs, étudiants, etc. est représenté par un objet. Ces objets forment une hiérarchie qui peut être manipulée et enrichie dynamiquement.

**2.3.** *Architectures pour l'adaptation de services*

Les nombreux systèmes qui supportent l'adaptation dynamique de services sont généralement constitués de trois parties :

–    La partie constituée du service qui peut être adapté dynamiquement selon des techniques très diverses [LED 02];

–    La partie chargée d'évaluer en continu l'ensemble constitué du service et de son contexte et qui effectue une tâche de monitoring ;

–    La partie de contrôle générant selon une logique qui lui est propre les ordres de reconfiguration.

Ces trois parties peuvent être physiquement distinctes. Dans certaines propositions, les trois parties sont présentes dans chaque composant. Par exemple, le modèle Molène proposée dans [SEG 00] permet à chaque composant de pouvoir être adapté, de posséder une partie liée à l'observation et éventuellement de générer des ordres de reconfiguration.

Dans [AKS 03] on propose une vue d'ensemble de la reconfiguration dynamique et des techniques d'adaptation. Dix techniques pour l'adaptation dynamique sont énumérées : composants insérables, algorithmes alternatifs - stratégies, programmation orienté aspect - AOP, filtres de composition, connecteurs changeables, patterns d'interaction décrits aussi dans [BLA 02], intergiciels réflexifs, injecteurs de comportement, interface adaptatives. Parmi ces techniques, une attention particulière est accordée aux intergiciels réflexifs [KON 02] comme OpenORB, dynamicTAO ou Xmiddle décrits dans [CAP 01]. Le principe d'une architecture réflexive est, par sa nature, adéquat à l'adaptation puisqu'il offre à la fois, la possibilité de comprendre, d'observer et de modifier l'architecture.

**2.4.** *Limitation des solutions existantes et notre position*

En général, les architectures adaptatives existantes utilisent pour la partie contrôle de l'adaptation un ensemble de règles spécifiques à chaque service pour décrire l'évolution d'un service en fonction de l'évolution du contexte. Ces propositions correspondent à un type de système fermé [DOW 01] dans lequel les évolutions sont généralement décrites avant la phase de déploiement. D'autres propositions, comme l'architecture Chisel complètent cette approche, [KEE 03]. Elles permettent à l'utilisateur, au développeur ou à l'administrateur de modifier et d'ajouter des nouvelles règles après le moment de la construction du service.

Nous considérons que la manière existante de réaliser la partie de contrôle de l'adaptation, à base des règles particulières à chaque service, a plusieurs inconvénients importants :

–    Le constructeur du service est obligé de prévoir au moment de la construction des règles pour tous les états visés du contexte. Il s'agit donc d'un effort important de la part du constructeur du service. Pour des contextes complexes et très variés, cette tache devient presque impossible à cause du nombre de règles à écrire.

–    Les règles sont spécifiques au service ainsi, pour chaque service il est nécessaire d'écrire écrire règles spécifiques.

–    Si le service doit s'adapter à de nouveaux contextes qui n'ont pas été identifiés lors moment de sa construction, l'intervention d'un programmeur est nécessaire pour qu'il définisse de nouvelles règles où réécrive le service.

–    La sémantique non explicitée de l'ensemble service-contexte est interprétable uniquement par le développeur du service. Il n'est donc pas possible de laisser la découverte des règles à la disposition d'un logiciel.

Nous avons fait le choix d'utiliser des intergiciels réflexifs comme support d'adaptation et de nous appuyer sur des formats existants pour représenter le contexte. Nous proposons une solution basée sur la description de chaque élément du service pour la partie de contrôle pour supprimer la description explicite de règles d'adaptation.

## 3. Proposition

### 3.1. *Objectif*

Notre objectif est de construire une architecture pour l'adaptation de services qui ne nécessite pas de règles particulières à chaque service. Si les approches existantes sont basées sur la construction des services doués dès leur construction des principes guidant leurs évolutions, nous essayons de d'adapter n'importe quel service par l'intermédiaire d'une architecture générale d'adaptation basée non pas sur une description des adaptations à réalisée mais sur une description précise de chaque élément du service.

### 3.2. *Architecture proposée*

L'architecture proposée (cf. figure 1) contient les trois parties précédemment identifiées d'un système adaptatif :

–    *Partie modifiable*. Cette partie est constituée par le service installé sur un intergiciel réflexif. Les éléments modifiables sont les composants et les différentes interconnexions entre ceux-ci.

–    *Partie monitoring*. Cette partie est représentée par des moniteurs qui observent les ressources et les profils utilisateurs. Ces éléments fournissent les données nécessaires à une description complète du service appelée méta-description de l'ensemble service-contexte.

– *Partie contrôle*. Cette partie est représentée par : l'Adaptateur qui, a partir de la description de l'ensemble service-contexte, décide la modification nécessaire pour adapter le service ; et d'un Assembleur qui exécute ce que l'Adaptateur lui dicte. L'Adaptateur utilise des composants existants trouvés dans la base de composants comme des solutions d'adaptation.

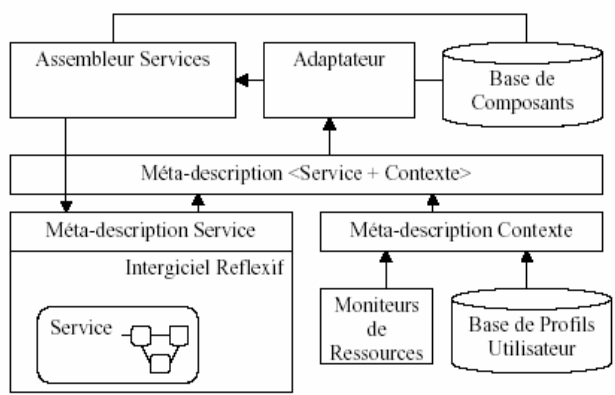

**Figure 1.** *Architecture pour l'adaptation dynamique de services*

### 3.2.1. *Exemple illustratif*

Un service de forum électronique permet aux étudiants étrangers à l'Université de partager des renseignements pratiques concernant leur installation sur le campus. Pour des raisons d'usage, la langue de discussion est l'anglais. Un étudiant étranger inscrit à l'Université souhaite lire et publier des messages sur ce forum.

### 3.2.2. *La méta-description de l'ensemble service-contexte*

La figure 2 présente une perspective sur la méta-description de l'ensemble service-contexte correspondante au scénario présenté dans la section 3.2.1. Nous trouvons trois plans différents : le plan physique contenant les éléments physiques du contexte, le plan des composants contenant l'ADL du service et les IDLs de composants et le plan de profils qui réunit le profil du service et le profil du contexte et qui indique le fonctionnement du service.

### 3.2.2.1. Le plan physique – éléments physiques du contexte

Le plan physique contient les éléments physiques du contexte. Dans ce cas il s'agit des utilisateurs du service de forum. Les éléments trouvés dans ce plan ont des projections dans le plan de profils. Ces projections forment la méta-description du contexte.

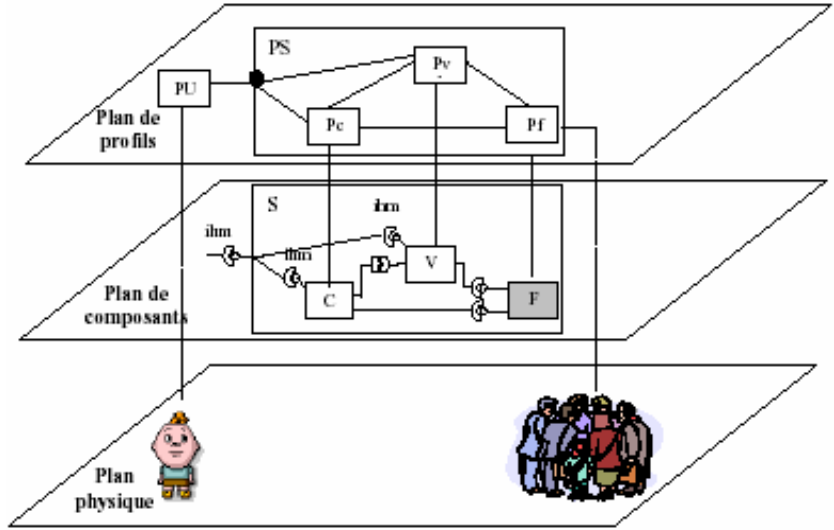

**Figure 2.** *La méta-description de l'ensemble service-contexte -  les différents plans de vue*

### 3.2.2.2. Le plan de composants – service, composants, assemblage

Le plan de composants contient le service S sous forme d'assemblage de composants logiciels. Pour notre exemple il est constitué : du composant V – Visualiser les messages publiés sur le forum, du composant C – Composer un nouveau message et du composant F – Forum qui contient tous les messages publiés sur le forum. L'IHM du service S est construite à partir d'une composition des IHM des composants C et V. Ce plan de composants représente ainsi la partie syntaxique de la méta-description du service.

Pour décrire le service dans ce plan nous utilisons un langage de type ADL qui décrit l'architecture interne du service, des descripteurs IDL et des connecteurs CCM pour les composants. Nous avons inclus aussi l'IHM parmi les connecteurs.

### 3.2.2.3. Le plan de profils – profil utilisateur, profil de composants, composition de profils

Le plan des profils est essentiel pour l'adaptation car il représente la partie sémantique de la méta-description de l'ensemble service-contexte. Pour l'exemple choisi il contient le profil du service - PS, et le profil utilisateur - Pu. PS est le résultat de la composition de profils de composants qui constituent le service : Pc, Pv, Pf. Ce plan contient la sémantique de l'ensemble service-contexte.

Le profil utilisateur contient, dans notre exemple, un seul paramètre : langue = 'FR' indiquant un utilisateur français. Le profil d'un composant indique comment ce composant fonctionne par rapport aux paramètres du contexte. Voici le profil pour un composant de traduction français – anglais :

```
profile>
        <component>TranslationFR_EN</component>
        <point>
                <interface>Translation</interface>
                <method>translate</method>
                <argument>text</argument>
                <argtype>String</argtype>
                <precondition>langue = 'FR' </precondition>
        </point>
        <point>
                <interface>Translation</interface>
                <method>translate</method>
                <returntype>String</returntype>
                <function>=</function>
                <postcondition>langue = 'EN' </postcondition>
        </point>
</profile>
```

Ce profil indique le comportement du composant par rapport à la langue : au niveau de l'interface « Translation » du composant « TranslationFR_EN » existe une pré-condition sur le paramètre « langue » qui demande la valeur « FR ». La valeur retourné aura la valeur « EN ». Ce composant modifie donc la langue et c'est grâce au profil que l'Adaptateur peut découvrir ce fait.

Le profil du service, qui est un composant composite, résulte par la composition des profils de composants C, V, F : F impose langue = 'EN' par convention, F est connecté avec C mais C est neutre de point de vue langue ainsi que la condition se propage au niveau de C. L'IHM de C est connecté à l'IHM de S ainsi que la condition se propage au niveau de l'IHM de S.

3.2.2.4. Les axiomes d'adaptation de l'ensemble service-contexte

Pour la partie sémantique de l'ensemble service-contexte, celle qui correspond au plan de profils, nous avons besoin de définir les axiomes qui sont vérifiés si un service est adapté à son contexte. Pour cet exemple il existe un seul axiome : nous avons besoin d'une adaptation de l'ensemble service-contexte si les valeurs des paramètres des profils sont contradictoires. Dans notre exemple, si l'utilisateur est français, il existe une contradiction pour le paramètre « langue », FR != EN au niveau de l'IHM du service S. Dans ce cas l'axiome d'adaptation n'est pas vérifié.

3.2.3. *L'Adaptateur – algorithme*

L'Adaptateur a le rôle d'adapter le service, dans ce but il doit effectuer plusieurs opérations :

— *Vérification*. Pour chaque paramètre des profils les axiomes d'adaptation sont vérifiés, s'ils sont tous respectés nous avons un service adapté.

— *Recherche d'une solution d'adaptation.* Si pour au moins un paramètre un axiome n'est pas respecté, il faut recherche une adaptation :

– Pour chaque paramètre qui ne vérifie pas les axiomes, l'Adaptateur construit un graphe qui a comme nœuds les interfaces de composants qui ont une relation avec ce paramètre et dont les arcs sont les interconnexions.

– Dans ce graphe l'adaptateur trouve les branches (enchaînement des nœuds) au niveau desquelles se trouvent des inégalités de valeur. Pour chaque branche l'Adaptateur cherche des composants qui, par insertion dans la branche, sont capables de rétablir l'égalité de valeurs.

— *Application de la solution.* Si l'adaptateur est arrivé à ce point, il peut appliquer la solution trouvée. Dans ce but, il peut soit prendre automatiquement la décision ou soit demander à l'utilisateur de choisir une solution. Dans notre exemple il va demander à l'utilisateur de décider s'il souhaite utiliser la traduction ou non, éventuellement en l'informant sur la qualité et le prix de cette traduction.

## 4. Implémentation

La figure 3 illustre le schéma du prototype réalisé.

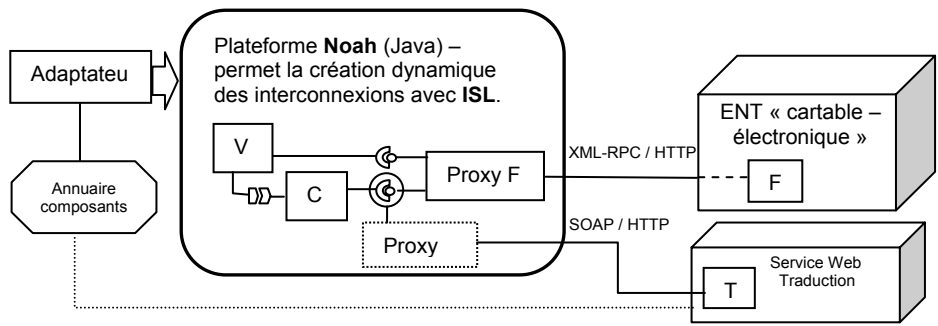

**Figure 3.** *Le schéma du prototype*

Nous pouvons observer dans cette figure que le service est constitué par les composants V – Visualiser forum, C – Composition message, Proxy F – un connecteur local vers le composant distant F – Forum. Ce service est extrait de l'environnement numérique de Travail (ENT) « Cartable électronique » [MAR 98]. Le composant T – Traduction, est ajouté dynamiquement sur l'initiative de l'Adaptateur, en utilisant le langage d'interactions ISL et la plateforme Noah [BLA 02]. L'Adaptateur trouve ce composant dans l'annuaire de composants, la recherche s'effectue à partir du profil du composant en respectant le typage des connexions.

La figure 4 illustre l'interface du prototype.

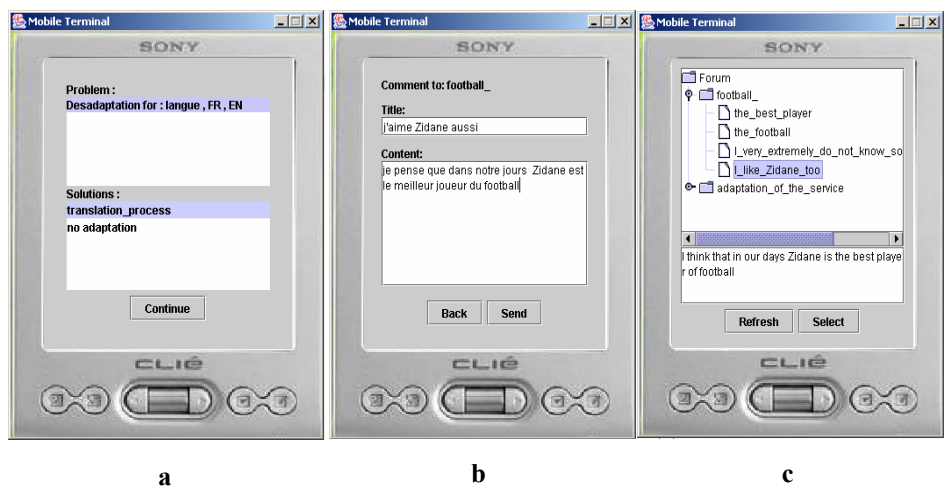

a                          b                          c

**Figure 4.** *Initialisation et utilisation du service de forum – captures d'écran*

Dans la figure 4.a, l'utilisateur choisit la solution car une inadaptation a été détectée. Dans les figures 4.b. et 4.c, le terminal permet à un utilisateur français d'écrire des messages dans sa langue naturelle et de les publier sur le forum en anglais. Si le même utilisateur écrit en italien ou en allemand, la langue est détectée et un composant de traduction correspondant sera inséré dynamiquement dans le service.

## 5. Conclusion

Dans cet article nous avons présenté une architecture qui permet l'adaptation dynamique de services. La différence entre notre proposition et les propositions existantes concerne essentiellement les aspects liés au contrôle de l'adaptation. Dans les solutions existantes le constructeur d'un service doit le doter dès sa conception de tous les composants nécessaires et des différentes règles d'adaptation. Dans notre proposition le contrôle est basé sur une méta-description de *l'ensemble service-contexte* dans lequel nous avons un ensemble limité des axiomes d'adaptation basé sur la sémantique du service. Il n'est pas nécessaire de décrire les différentes règles d'évolution, celles-ci seront découvertes par analyse des cas d'inadaptation sous la seule réserve que chaque composant soit décrit avec son profil.

Pour prouver le fonctionnement de cette architecture nous avons implémenté un prototype de service de forum. Ce service est crée initialement pour des utilisateurs anglophones mais l'architecture proposée permet d'adapter ce service par ajout dynamique d'un composant de traduction si la langue de l'utilisateur n'est pas l'anglais.

Le principal désavantage de notre l'architecture est sa complexité. Il reste encore à généraliser et simplifier le modèle afin que les différents paramètres de profils se composent plus simplement tout en gardant une richesse nécessaire à l'expression des axiomes d'adaptation.

Malgré ces limitations et difficultés, nous considérons que l'avenir de l'adaptation appartient à la composition sémantique des services. Pour réaliser cela, le fonctionnement de l'ensemble service-contexte doit être compréhensible aussi pour la machine et pas seulement pour le constructeur humain de services.

## 7. Bibliographie